**Norman Yao** is an assistant professor in the physics department at the University of California, Berkeley. **Chetan Nayak** is the director of Microsoft Station Q in Santa Barbara, California, and a professor in the physics department at the University of California, Santa Barbara.

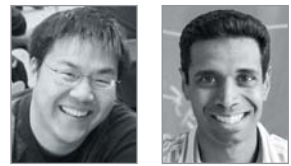

Norman Y. Yao and Chetan Nayak

When the discrete time-translation symmetry of isolated, periodically driven systems is spontaneously broken, a new phase of matter can emerge.

Time is an outlier. Although relativity attempts to unify time and space into one seamless object, time is still special in many contexts. One manifestation of that special nature is the difference between time-translation symmetry and other symmetries. The spatial translational symmetry of atoms, the rotational symmetry of spins, and many others can be spontaneously broken. And they are.

When a symmetry is spontaneously broken, a system becomes less symmetrical than its parent Hamiltonian. A crystalline solid is a classic example. Outside the crystal, interactions between atoms are the same anywhere in space—they are continuously translationally invariant. But a crystal's ground state has a preferred set of lattice points, picked out from a family of energetically equivalent choices by even an infinitesimal perturbation, and the crystal is invariant only if shifted by specific amounts.

Spontaneous symmetry breaking is a unifying concept in modern physics. Examples abound, including magnets, superconductors, and, according to the standard model of particle physics, the whole universe: According to an intuitive picture of the Higgs mechanism, spontaneous symmetry breaking underlies the origin of particle masses (see PHYSICS TODAY, September 2012, page 14). That ubiquity seems to suggest that almost any symmetry can be broken.

But for most people, time-translation symmetry—in which a system's governing equations are unchanged by going to earlier or later times—somehow feels different. Schrödinger's equation dictates that a system's ground state, and indeed any energy eigenstate, must transform trivially under time translation and pick up only a simple overall phase factor. Examples abound of crystalline solids, which are periodic in space, but "time crystals," so named by Frank Wilczek,[1] are a mere fantasy.[2] Or so we thought.

Recently, however, physicists have realized that in periodically driven closed quantum systems, *discrete* time-translation symmetry isn't actually so different from other symmetries.[2] It can be spontaneously broken[3–6] and can protect topological states of matter in a manner completely analogous to other symmetries. The epiphany that discrete time-translation symmetry can be treated on par with other, more conventional symmetries has revised our understanding of time and even has had an almost immediate effect on experiments.[6–8]

The emergent properties of strongly interacting, periodically driven many-particle systems have led to the concept of a Floquet time crystal or discrete time crystal: a state of matter that exhibits spontaneously broken discrete time-translation symmetry;[3] we unpack the definition of that technical term below. Such a state was assumed to be impossible, partly because of unambiguous proofs that rule out the breaking of continuous time-translation symmetry in equilibrium systems.[2,9] But the proofs leave the door open to the breaking of discrete time-translation symmetry in inherently nonequilibrium contexts, and Floquet time crystals serve as an ideal example.

Floquet time crystals derive their name from French mathematician Gaston Floquet (1847–1920), who studied ordinary differential equations with periodic time dependence. Although Floquet time crystals are outside of equilibrium, in some sense they represent the mildest sort of nonequilibrium system: In a time-dependent rotating basis, they are actually equivalent to equilibrium systems, a feature we dub "cryptoequilibrium" to underscore that hidden equilibrium nature.

A precise definition of discrete time-translation symmetry breaking (TTSB) leads to smoking-gun experimental signatures of a Floquet time crystal. Moreover, it enables one to tease apart the subtle features of a discrete time crystal and to draw sharp distinctions with a host of superficially similar-looking but quite distinct nonequilibrium phenomena,[10] some of which date back a century or more (see the article by Ray Goldstein on page 32).

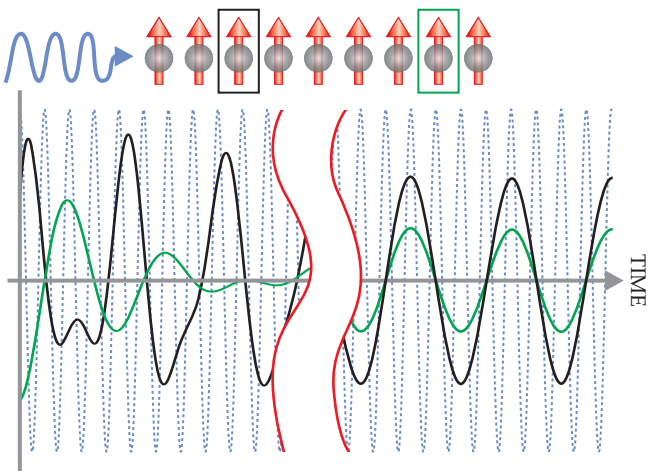

**FIGURE 1. A PERIODICALLY DRIVEN SYSTEM.** In this schematic of a chain of spins, we depict how two of the spins (black and green) respond when driven by an oscillating external source (dashed blue line). After some initial transient behavior, the spins fall into lockstep at a frequency that is ¼ of the drive frequency. Such a subharmonic response is characteristic of a discrete time crystal: Its periodicity breaks the discrete time-translation symmetry of the drive.

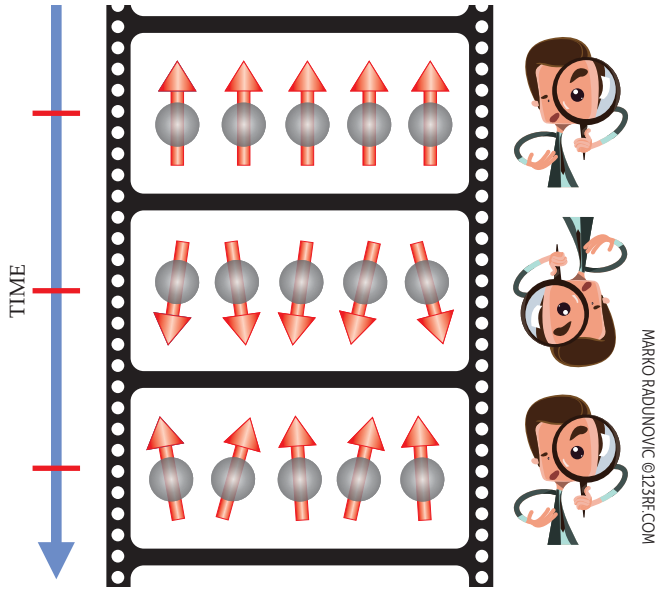

**FIGURE 2. CRYPTO-EQUILIBRIUM IN A DISCRETE TIME CRYSTAL.** In a $2T$-periodic time crystal, the spin system's overall orientation flips during each driving period, so it takes two periods for the spins to return to something resembling their initial state. But to someone viewing the system at fixed intervals (that is, stroboscopically) from a frame of reference that flips with the spins, the system appears to be in equilibrium, and it exhibits many features of an equilibrium system; no entropy is generated and oscillations need no other inputs to sustain them. We therefore say the system is in crypto-equilibrium.

## What is a discrete time crystal—and what isn't

Discrete time-translation symmetry breaking manifests itself in three key ways:
▶ Broken symmetry: After a possible initial transient period, the system exhibits late-time oscillations with a period longer than that of the drive.
▶ Crypto-equilibrium: No entropy is generated by the late-time oscillations.
▶ Rigid long-range order: The oscillations remain in phase over arbitrarily long distances and times.

Making those notions more precise is a bit subtle. When talking about states of matter, one typically starts with a preferred state, the ground state, but not in discrete time crystals: Periodically driven systems do not have a ground state. Again, a comparison to spatial crystals helps. Consider a Floquet system that's driven at period $T$, so that for any time $t$ the Hamiltonian $H(t)$ satisfies $H(t+T)=H(t)$. The discrete spatial-translation symmetry of a one-dimensional spatial crystal leads to electron states having a quasi-momentum that is only defined modulo $2\pi/a$, where $a$ is the period of the crystal lattice. Analogously, the discrete time-translation symmetry of a periodically driven system leads to eigenstates having a quasi-energy that is only defined modulo $\Omega \equiv 2\pi/T$, and there is no preferred state with a minimum value of the quasi-energy. Thus any definition of TTSB in periodically driven, nonequilibrium systems cannot be cast in terms of ground-state or low-energy properties.

But if we don't restrict ourselves to ground-state properties, then oscillations with a frequency $\omega \neq \Omega$—a requirement for TTSB—can be realized fairly easily. Even for a simple harmonic oscillator, if our initial state is a superposition of two eigenstates then the system will naturally exhibit oscillations at a frequency given by the difference between the eigenstate energies. But in most systems, a generic initial state will not lead to late-time oscillations with $\omega \neq \Omega$. So any good definition of TTSB in Floquet systems must generalize the ground-state or equilibrium notion of spontaneous symmetry breaking in such a way that the oscillatory behavior does not depend on the choice of initial state.

That requirement—and thus a precise definition of TTSB in periodically driven systems—can be stated in a remarkably compact form: A discrete time crystal is a state of matter in which the Floquet eigenstates are necessarily "cat states," that is, entangled superpositions of macroscopically distinct states. One immediate corollary is that because any initial physical state we can prepare must be a superposition of such Floquet cat states, *all* preparable initial states will exhibit oscillations.

Let us further unpack that definition. Since the Floquet eigenstates of a discrete time crystal aren't preparable, we can readily distinguish the discrete time crystal from, for example, a simple harmonic oscillator. As mentioned above, whether the harmonic oscillator displays time-periodic behavior depends strongly on the choice of initial state one prepares, and most importantly, nothing prohibits us from preparing a harmonic oscillator in its ground state. By stark contrast, in a discrete time crystal every physically preparable initial state will exhibit oscillations at late times.

The key features of discrete time crystals explicitly distinguish them from a multitude of other systems (see the table on page 46) that exhibit oscillations with unexpected periods.[10]

## AN ANALOGY: METHANE ON GRAPHITE

The features that make discrete time crystals truly special can be hard to fully appreciate. To illustrate the essence of a discrete time crystal, we turn to a spatial analogy based on the surface of graphite, the familiar hexagonal crystalline allotrope of carbon. Using that analogy, we will explore three key concepts: the breaking of a discrete translational symmetry, long-range spatial and temporal ordering, and the importance of many-body interactions for stabilizing the broken symmetry.

**Discrete translational symmetry.** Due to the underlying arrangement of the carbon atoms, the surface of graphite breaks the continuous translational symmetry of space into the discrete spatial translational symmetry of a honeycomb lattice (gray in the figure). Suppose that methane molecules (yellow) are now adsorbed on the surface. At temperatures greater than approximately 60 K, the methane molecules form a two-dimensional liquid on the graphite surface. As the system is cooled down, the methane solidifies.

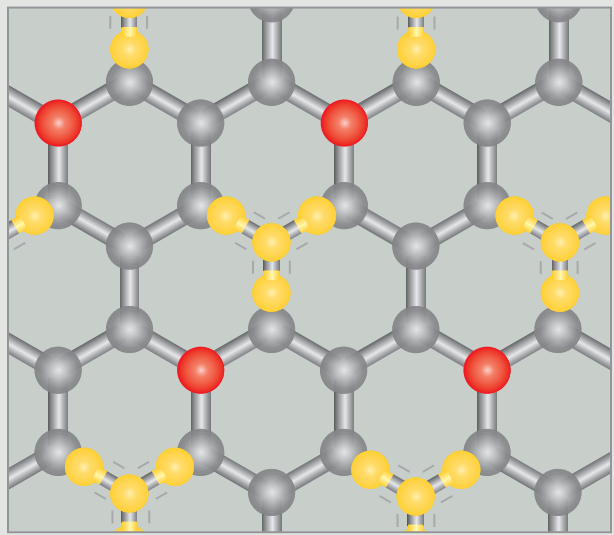

The solidified methane molecules can't match the underlying graphite lattice. Depending on its density, the methane may crystallize into a preferred sublattice—the one occupied in the figure, perhaps, or the alternative sublattice indicated in red. In spontaneously choosing a sublattice, the methane adopts an arrangement of lower symmetry than the graphite's honeycomb lattice: It breaks an already discrete spatial translational symmetry.

Now consider the analogue in time. Driving a system periodically establishes discrete time-translational symmetry. Much like graphite's repeating pattern of carbon atoms in space, periodic driving leads to a repeating pattern in time: The system's Hamiltonian returns to itself after every full driving period. The breaking of that discrete time-translation symmetry would manifest as the system's behavior "crystallizing" on a "sublattice" in time; the prime example is an observable whose response has double the period of the underlying drive.

**Long-range order.** If we take a snapshot of a portion of the graphite's surface, the state of the methane molecules—liquid or crystalline—may not be obvious. On one hand, fluctuations in the crystalline state can take methane molecules from their preferred sublattice to other sublattices, and make the crystalline state appear liquid. On the other hand, a small region of the liquid state may momentarily look crystalline; as anyone who has done a belly flop in a swimming pool can attest, water seems rather solid on short time scales.

But if we look over a large enough region of the graphite's surface or on long enough time scales, the liquid's momentary crystallinity will wash out. For large regions, sufficiently separated crystalline patches will choose different sublattices; for long times, a small crystalline cluster on one sublattice is equally likely to be on a different sublattice at a later snapshot.

Therein lies the essence of long-range order: It distinguishes symmetry-broken (crystalline) and unbroken (liquid) states. To diagnose the crystalline state of the adsorbed methane molecules, it is crucial to make sure that the molecules prefer the same sublattice in distant regions and at different times—that they have long-range spatial and temporal ordering. The same must hold true for the time crystal.

**Role of interactions.** The long-range order that characterizes the symmetry-broken crystalline state of adsorbed methane molecules requires the presence of interactions: Only through the repulsion between one methane molecule and its neighbor, and then between that neighbor and its next neighbor, can the entire system manage to ensure that all the molecules prefer the same sublattice. If a fluctuation puts the molecules in some region onto the wrong sublattice, the resulting sublattice mismatch creates a domain wall, which costs energy. If the system is cold enough, that energy cost will cause subsequent fluctuations to put the misaligned region back onto the preferred sublattice.

That realignment critically relies on the presence of interactions. Thus one expects that nontrivial time-crystalline order should not result from the dynamics of individual particles but rather from the collective synchronization of many strongly interacting degrees of freedom.

---

The box above offers some insights into those features through a spatial analogue: methane adsorbed on the surface of graphite.

In principle, an oscillation with any frequency $\omega \neq \Omega$ would break the discrete time-translation symmetry. However, we will focus on the subharmonic case, $\omega = \Omega/n$ for some integer $n > 1$; figure 1 depicts such a system with $n = 4$.

Suppose that we measure a nonequilibrium Floquet system stroboscopically—that is, at regular fixed intervals, like frames of a movie reel. When viewed at multiples of the drive period, $t = kT$ for $k = 1, 2, \ldots$, then as depicted in figure 2, there is some time-dependent frame of reference in which we will be unable to tell that the system is not in static equilibrium. Instead of being observed in the system, the oscillations are subsumed into the frame's time dependence.

We say that a periodically driven system is in crypto-equilibrium if there exists some reference frame, possibly time dependent, in which the system is indistinguishable from a system in thermal equilibrium, if measured stroboscopically. (In fact, if we measure the system at $t = knT$ for positive integers $k$, then there's no need to go into a moving frame at all; even in the fixed lab frame, the system looks like an equilibrium system.) In many of the cases we discuss below, the appropriate frame is similar to the rotating frame that is routinely used to simplify the analysis of NMR experiments. For a discrete time crystal to be in crypto-equilibrium, however, the rotating frame must play a more powerful role than just simplifying the analysis: It must transform the periodically driven system into a stationary, equilibrium one.

Crypto-equilibrium requires that the periodic drive add no entropy, and it enables a discrete time crystal to exhibit rigid long-range order. By contrast, oscillating chemical reactions and convection are inescapably nonequilibrium: They are irreversible processes that generate entropy and require a constant flux of incoming reactants or a maintained thermal gradient to drive the oscillatory behavior. While the oscillations appear periodic at first sight, they generally fall out of lockstep when probed over longer times and at distant locations. That behavior distinguishes discrete time crystals from other oscillatory nonequilibrium phenomena as a matter of principle.

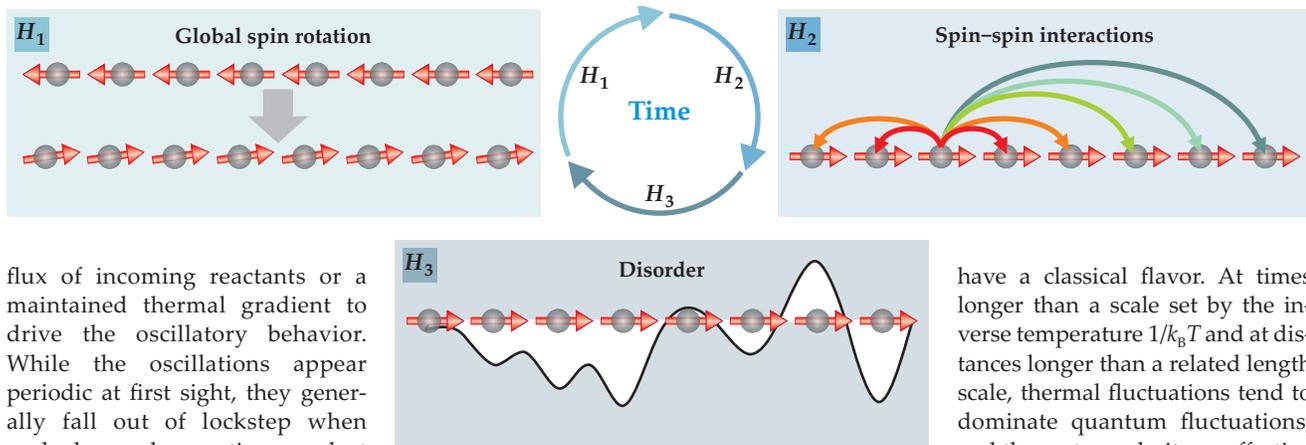

FIGURE 3. A SIMPLE SPIN MODEL that cycles through three stages: approximate spin flips ($H_1$), strong spin–spin interactions ($H_2$), and disorder ($H_3$). That three-step evolution captures the key features that support discrete time crystals. (Adapted from ref. 7.)

The long-range order of discrete time crystals is a remarkable phenomenon. Since they are not in thermal equilibrium, their characteristic rigidity is not reliant on low energy or temperature; it depends instead on emergent features that control the strength of energy and quantum fluctuations. To fully appreciate those features, we must delve deeper into the physics of isolated, periodically driven quantum systems.

## Thermalization and its breakdown

That periodically driven systems can be in crypto-equilibrium is rather counterintuitive: A generic periodically driven, isolated system will absorb energy until it looks, locally, like an infinite-temperature state.[11] If that were the full story, it would preclude time crystals, since Floquet systems would all simply end up as featureless infinite-temperature states in the thermodynamic limit. Luckily, there are two situations (at least) in which that discouraging conclusion is not correct.

First, if the drive frequency is very large compared with the local energy scales of the system, then the system can only absorb energy from the drive by spreading it out over many excitations.[12–14] Consequently, heating occurs very slowly, and there is a long-lived quasi-steady state—a so-called "pre-thermal" state—in which ordered phases of matter can occur.

Second, if a system has a high density of frozen-in impurities, then a phenomenon called many-body localization can occur.[15] MBL prevents the spreading of energy because the impurities trap excitations, even highly excited ones. For periodically driven Floquet–MBL systems, the energy spreading and thermalization necessary for drive-induced heating cannot occur, so nonequilibrium ordered states of matter can survive indefinitely. Those two loopholes—pre-thermal states and MBL—both basically rely on the discreteness of quantum mechanical energy levels, and they open the door for stabilizing various driven systems, including discrete time crystals.

The quantum mechanical nature of pre-thermal states and MBL raises a natural question: Does time-crystalline order require quantum mechanics? In general, broken-symmetry states have a classical flavor. At times longer than a scale set by the inverse temperature $1/k_B T$ and at distances longer than a related length scale, thermal fluctuations tend to dominate quantum fluctuations, and the system admits an effective classical description. One can thus ask about the specific role played by quantum mechanics. Is it only important insofar as it prevents the system from heating up? Or is quantum mechanics essential for enabling the subharmonic synchronization characteristic of a discrete time crystal?

Recent work suggests the former. A dissipative, coupled chain of classical nonlinear pendula, it turns out, can exhibit a phase transition between a time-crystalline phase and a phase with unbroken symmetry.[16] That suggests a third strategy for suppressing the order-ravaging, drive-induced heating of Floquet systems: coupling to a dissipative bath.[3] The system would not be in crypto-equilibrium, since entropy is created in the bath. But when the drive is noisy, such a system may be smoothly connected in some precise way to a crypto-equilibrium system. It is natural to ask, given their departure from equilibrium, whether there could be time crystals that break continuous time-translation symmetry. In the pre-thermal regime, the answer is in fact yes: Their experimental signatures have been observed[17] and recently clarified.

## Models . . .

A simple set of spin models can host discrete time crystals.[3–6] In particular, consider a Floquet Hamiltonian with $H(t+T)=H(t)$ and total evolution time $T=t_1+t_2+t_3$:

$$H(t) = \begin{cases} H_1 & \text{for a time } t_1 \\ H_2 & \text{for a time } t_2 \\ H_3 & \text{for a time } t_3. \end{cases}$$

Here, $H_1$, $H_2$, and $H_3$, illustrated in figure 3, are time-independent Hamiltonians given by

$$H_1 = \frac{\pi}{2t_1} g \sum_i \sigma_i^x, \quad [\text{Global spin rotation}]$$

$$H_2 = -\sum_{\langle i,j \rangle} J_{ij} \sigma_i^z \sigma_j^z, \quad [\text{Spin–spin interactions}]$$

$$H_3 = -\sum_i \left( h_i^x \sigma_i^x + h_i^y \sigma_i^y + h_i^z \sigma_i^z \right), \quad [\text{Disorder}]$$

where $\sigma_i^x$, $\sigma_i^y$, and $\sigma_i^z$ represent the Pauli spin operators for spin $i$. For a 1D system, many-body localization occurs, and this model supports a period-$2T$ discrete time crystal provided that the local magnetic field vector $\mathbf{h}_i$ is small and there is sufficient disorder. In more than one dimension, pre-thermal states occur if the energy scales of $H_2$ and $H_3$ are small compared to the drive frequency $\omega$, and if $g \approx 1$ there will again be a period-$2T$ discrete time-crystal phase for a finite fraction of possible initial states. Similar, slightly more complicated models admit period-$nT$ discrete time crystals; the simplest of them feature an $n$-level system with clock-like transitions between the levels.[8]

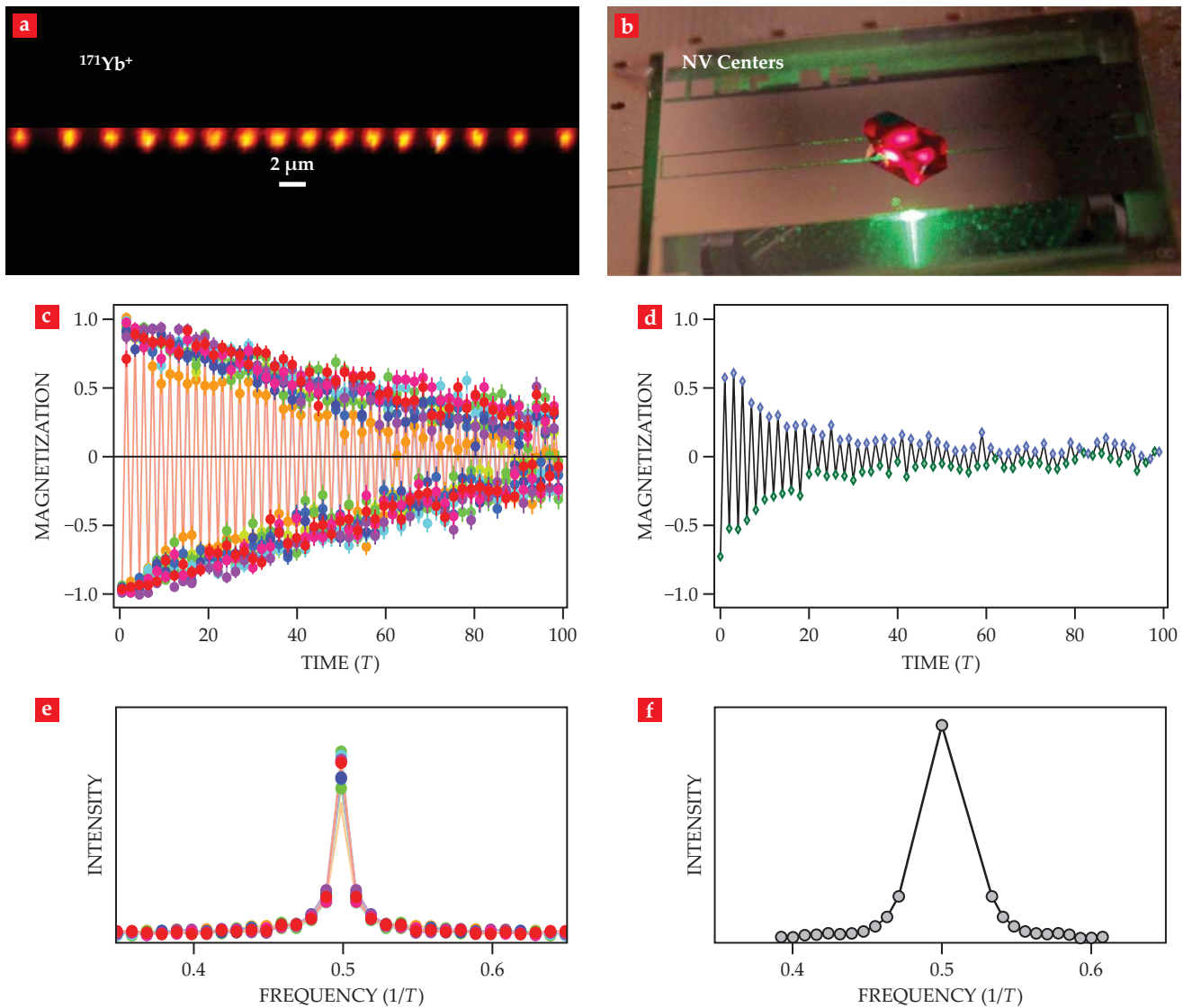

**FIGURE 4. EXPERIMENTAL DEMONSTRATIONS.** The first signatures of discrete time-crystalline order were reported in two spin systems. **(a)** A one-dimensional chain of trapped ytterbium ions.[7] Each ion had an effective spin-½ state created from two of its hyperfine sublevels, and ion–ion interactions generated a lattice arrangement. **(b)** A 3D ensemble of nitrogen–vacancy defects (NV centers) in diamond.[8] The NV centers fluoresce red under green laser illumination. **(c, d)** Each system was driven by a Floquet sequence, like that in figure 3, for about 100 cycles. **(e, f)** Fourier transforms of the measured magnetizations show sharp oscillations at half the cycle frequency $1/T$, where $T$ is the cycle period. (Adapted from refs. 7 and 8.)

The above model has a few simple limits. For $g=1$ and $J=h^x=h^y=0$, the first part of the Floquet sequence applies an exact $\pi$-pulse, which flips all the spins, and the third part conserves the $z$-component of each spin. As a result, the spins are reversed after one period and return to their initial configuration after two. That behavior is completely trivial, though, since the spins do not interact with one another. Consequently, much like the period doubling of individual nonlinear oscillators,[10] the behavior does not survive the addition of perturbations and fluctuations.

Now suppose we turn on a finite spin–spin coupling $J$. $H_1$ still applies an exact $\pi$-pulse, and $H_3$ conserves the spins' $z$-components. But the net behavior is no longer trivial—two key differences emerge on closer inspection. First, the Floquet eigenstates are now superpositions of two states that differ by a flip of all spins; hence, they are macroscopic cat states. Second, and most remarkably, the period-doubled behavior of the system is qualitatively unchanged by small perturbations such as $g \neq 1$. That is precisely a manifestation of the rigidity that we

| Time crystals' defining traits | Time crystal | Period-doubled nonlinear dynamical system | Mode-locked laser | Parametric down-conversion | NMR spin echo | Belousov–Zhabotinsky reaction | Convection cells | AC Josephson effect |
|---|---|---|---|---|---|---|---|---|
| Many-body interactions | ✓ | X | ✓ | ✓ | X | X | ✓ | ✓ |
| Long-range order | ✓ | X | X | X | X | X | X | X |
| Crypto-equilibrium | ✓ | X | X | X | X | X | X | X |

▶ **Time crystals** have many-body interactions that establish long-range order in both space and time. Long-range order renders the systems stable against small perturbations that respect the discrete time-translation symmetry of an ideal, perfectly periodic drive. And time crystals exhibit so-called crypto-equilibrium: They don't require a sustaining external flux and don't generate entropy; indeed, in a rotating frame of reference they appear to be in equilibrium.

▶ An oscillating **nonlinear dynamical system** can exhibit **period-doubling**, but the oscillations do not remain in phase over arbitrarily long times—unless fluctuations are dissipated by coupling to a bath and thereby generate entropy.

▶ **Mode-locked lasing** is a many-body effect, but fluctuations in cavity length will destroy long-term phase coherence.

▶ **Parametric down-conversion**, in which one photon will split into two, is a stochastic process that lacks even short-time order.

▶ **NMR spin echoes** cause oscillating transverse magnetization, but the magnetization must be constantly regenerated and is sensitive to perturbations in pulse length and field strength.

▶ **Belousov–Zhabotinsky** chemical reactions oscillate, but they lack long-range spatial order, require an external flux of reactants, and generate entropy.

▶ **Convection cells** occur when a fluid heated from below rises, eventually cools, and sinks. But the cells lack long-range spatial and temporal order, require an external gradient, and generate entropy.

▶ The **AC Josephson effect** arises when a voltage applied across a superconducting tunnel junction induces an oscillating supercurrent. However, phase slips cause dissipation and thereby generate entropy.

Extended systems of coupled Josephson junctions or nonlinear oscillators can, through their coupling, stabilize oscillations with many of the features of time crystals.[16]

---

expect of an ordered, broken-symmetry phase. Although the perturbations lead to fluctuations, those fluctuations do not cause the discrete time crystal's oscillations to fall out of phase at distant locations or distant times. The system is long-range ordered!

## . . . and experiments

Two essential features unify spin models that support discrete time crystals: strong interactions and an effective magnetic field that implements an approximate but not necessarily exact $\pi$-pulse. Those two attributes are realizable in a wide range of quantum optical systems ranging from trapped ions and solid-state spin defects to Rydberg atoms and polar molecules. The signatures of time-crystalline order were reported last year in two wildly disparate platforms: a 1D array of trapped ytterbium ions[7] (figure 4a) and a 3D ensemble of nitrogen–vacancy spin defects (NV centers) in diamond (figure 4b).[8] We consider each in turn.

▶ **Trapped ions.** Chris Monroe's group at the University of Maryland[7] worked with 1D chains of up to 14 $^{171}$Yb$^+$ ions. The interplay between the trapping forces and the ions' natural Coulomb repulsion produced the (spatial) crystalline configuration seen in figure 4a. For each ion, the researchers constructed from two hyperfine states an effective spin-½ degree of freedom that could be optically prepared, manipulated, and observed.

The researchers repeatedly cycled their ion chain through the three-Hamiltonian sequence of figure 3. One pair of lasers generated approximate $\pi$-pulses to flip the spins. A second set of lasers coupled the ions' hyperfine spin states to the motional vibrations of the crystalline configuration and produced long-range spin–spin interactions that fell off as a tunable power law. Lastly, a tightly focused laser beam that was scanned along the ion chain addressed each spin in turn and generated a site-dependent, disordered longitudinal field.

Long-range interactions, however, tend to disfavor the MBL needed to realize time-crystalline behavior. So the researchers worked in a parameter regime in which, according to numerical simulations, the disorder is strong enough and the long-range power law weak enough that the system could exist in the putative MBL phase.

With the ions' Floquet evolution in hand, the experimenters observed two of the key features predicted for time-crystalline order: interaction-stabilized period doubling (figure 4e) and a phase transition to a non-time-crystalline state as $\pi$-pulse deviations increased.

▶ **Nitrogen–vacancy centers.** Mikhail Lukin's group at Harvard University observed signatures of time-crystalline order in ensembles of NV centers.[8] Each NV center behaved as a spin-1 magnetic impurity in the surrounding diamond lattice (figure 4b). By applying a small magnetic field along the NV center's axis, the researchers lifted the degeneracy of the $m_s = \pm 1$ states and isolated an effective two-level spin system. The resulting spin system could be initialized, manipulated, and detected using a combination of optical and microwave radiation. The NV centers interacted with one another via long-range magnetic dipole–dipole interactions, and disorder was naturally present in the system because of other randomly distributed paramagnetic impurities inside the lattice.

Those features all seem quite reminiscent of the trapped-ion system, and the systems' signatures of time-crystalline order were similar. But essential differences exist between the trapped-ion and NV platforms, which makes the similarity of their observations all the more surprising. In particular, NV systems are 3D, so many-body localization is not expected. Moreover, it does not appear that pre-thermal states occur in NV systems, either; the energy of the initial state is too high. Yet within a comparable number of Floquet evolution cycles, the NV experiments—like the trapped ions—demonstrated that interactions help stabilize the subharmonic Fourier response of a discrete time crystal (figure 4f). Even more recently, Sean Barrett's group at Yale University has reported similar observations in NMR experiments focusing on the spin-½ phosphorus nuclei in ammonium dihydrogen phosphate (ADP).[18]

Whereas the trapped-ion experiments represent a more canonical model of a discrete time crystal, the NV and ADP ex-periments suggest the possibility that a broader class of time-crystalline behavior can be observed, so long as a mechanism exists for slowing down thermalization and heating. It may be that the interplay between dimensionality and dipolar interactions inherently results in a relatively slow approach to thermal equilibrium and that the observed time-crystalline order is merely manifest in that transient regime.

## Outlook

The recent experiments present preliminary evidence for a discrete time crystal. But improved coherence times are needed if experiments are to truly demonstrate long-range temporal ordering—that is, the observed spin oscillations remain in phase over extended times.

The discrete time crystals discussed here are not the only phases of matter of periodically driven systems. Such systems can have other phases, including so-called topological phases and symmetry-protected topological (SPT) phases. The most remarkable of them share a property with discrete time crystals: They exhibit nonstationary behavior, even when observed stroboscopically. That behavior is more subtle in topological or SPT phases, however, than in discrete time crystals.

Unlike topological and SPT phases, discrete time crystals have the additional property that they are genuine spontaneous symmetry-breaking phases. Moreover, the symmetry that they break, time-translation symmetry, cannot be broken in a stable phase in thermal equilibrium. Thus discrete time crystals lie at the intersection between nonequilibrium and exotic spontaneous symmetry-breaking phases of matter. Although it is too early to say, the greatest long-term impact of time crystals may well be that they have opened our eyes to the new world of nonequilibrium phases of matter.

## REFERENCES

1. F. Wilczek, *Phys. Rev. Lett.* **109**, 160401 (2012).
2. H. Watanabe, M. Oshikawa, *Phys. Rev. Lett.* **114**, 251603 (2015).
3. D. V. Else, B. Bauer, C. Nayak, *Phys. Rev. Lett.* **117**, 090402 (2016); *Phys. Rev. X* **7**, 011026 (2017).
4. V. Khemani et al., *Phys. Rev. Lett.* **116**, 250401 (2016).
5. C. W. von Keyserlingk, S. L. Sondhi, *Phys. Rev. B* **93**, 245146 (2016).
6. N. Y. Yao et al., *Phys. Rev. Lett.* **118**, 030401 (2017); erratum **118**, 269901 (2017).
7. J. Zhang et al., *Nature* **543**, 217 (2017).
8. S. Choi et al., *Nature* **543**, 221 (2017).
9. P. Bruno, *Phys. Rev. Lett.* **111**, 070402 (2013).
10. For a discussion of many of these systems, see S. H. Strogatz, *Nonlinear Dynamics and Chaos: With Applications to Physics, Biology, Chemistry, and Engineering*, 2nd ed., CRC Press (2015).
11. L. D'Alessio, M. Rigol, *Phys. Rev. X* **4**, 041048 (2014).
12. T. Kuwahara, T. Mori, K. Saito, *Ann. Phys.* **367**, 96 (2016).
13. D. Abanin et al., *Commun. Math. Phys.* **354**, 809 (2017).
14. F. Machado et al., http://arxiv.org/abs/1708.01620.
15. R. Nandkishore, D. A. Huse, *Annu. Rev. Condens. Matter Phys.* **6**, 15 (2015).
16. N. Y. Yao et al., http://arxiv.org/abs/1801.02628.
17. C. Urbina, J. F. Jacquinot, M. Goldman, *Phys. Rev. Lett.* **48**, 206 (1982).
18. J. Rovny, R. L. Blum, S. E. Barrett, http://arxiv.org/abs/1802.00126.   **PT**